\documentclass[preprint,authoryear,12pt]{elsarticle}
\makeatletter
\def\ps@pprintTitle{%
 \let\@oddhead\@empty
 \let\@evenhead\@empty
 \def\@oddfoot{}%
 \let\@evenfoot\@oddfoot}
\makeatother
\usepackage{graphicx}
\usepackage{url}
\usepackage{amssymb}
\journal{New Astronomy}

\begin{document}

\begin{frontmatter}

\title{A Gaia based analysis of open cluster Berkeley 27}

\author[label1]{Devesh P. Sariya}
\ead{deveshpath@gmail.com}
\author[label1]{Ing-Guey Jiang}
\author[label2,label3]{D. Bisht}
\author[label4]{R. K. S. Yadav}
\author[label2]{G. Rangwal}

\address[label1]{Department of Physics and Institute of Astronomy, National Tsing-Hua University, Hsin-Chu, Taiwan}
\address[label2]{Indian Institute of Astrophysics, Koramangala II Block, Bangalore, 560034, India}
\address[label3]{Indian Center for Space Physics, 43 Chalantika, Garia Station Rd., Kolkata 700084, India}
\address[label4]{Aryabhatta Research Institute of Observational Sciences, Manora Peak, Nainital 263002, India}

\begin{abstract}
In this paper, we have used the Gaia's Data Release-3 (DR3) data
to study an intermediate-age open cluster Berkeley 27 (Be 27).
A total of 131 most probable cluster members are picked
within the cluster's radius based on the membership probability ($> 80\%$).
The cluster's radius was estimated as 3.74 arcmin.
The mean proper motion (PM) of Be 27 was determined
to be ($\mu_{\alpha} cos{\delta}$, $\mu_{\delta}$)=
($-1.076\pm0.008$, $0.152\pm0.007$)~mas~yr$^{-1}$.
The blue straggler stars (BSS) of the cluster were found to be
located in the central region.
Theoretical isochrones of metallicity Z$_{metal}$= 0.008 were compared to the
color-magnitude diagram (CMD) of Be 27.
As a result, a heliocentric distance of 4.8$\pm$0.2 kpc
and log (age) = 9.36$\pm$0.03 were determined for Be 27.
The Galactic orbits are derived using the Galactic potential model
which demonstrate that Be 27 follows a circular path around the Galactic center.
The cluster does not seem to be affected much by the tidal forces 
from the Galactic thin disk.

\end{abstract}


\end{frontmatter}

\section{INTRODUCTION}
\label{intro}

Open clusters are the abundant star clusters in the Galactic disk region. 
The open clusters have a vast age range from a few million years to a few
billion years. Due to their age range, the open clusters are used in studying 
the star formation process and stellar evolution 
(Lada \& Lada 2003, Kim et al. 2017).
The component stars of an open cluster have similar 
age, heliocentric distance, metal content, etc. (Yadav et al. 2011). 
Stars of a cluster have a similar mean motion (Sariya et al. 2021a).
Therefore, with the knowledge of the proper motion (PM) components 
of individual stars in the direction of a cluster, 
their membership status can be defined. 
Such a study provides a refined sample of cluster stars 
for the derivation of the basic parameters 
(Cudworth 1997; Joshi et al. 2016; Bisht et al. 2020,2021; Sariya et al. 2021a,b). 
Gaia satellite data has provided several essential results
for our Galaxy and its constituents 
(e.g., Cantat-Gaudin et al. 2018; Soubiran et al. 2018; 
Gao 2018; Castro-Ginard et al. 2019; Ding et al. 2021;
Penoyre et al. 2022; Belokurov et al. 2022).

\begin{figure*}
\begin{center}
\includegraphics[width=14.5cm, height=9.5cm]{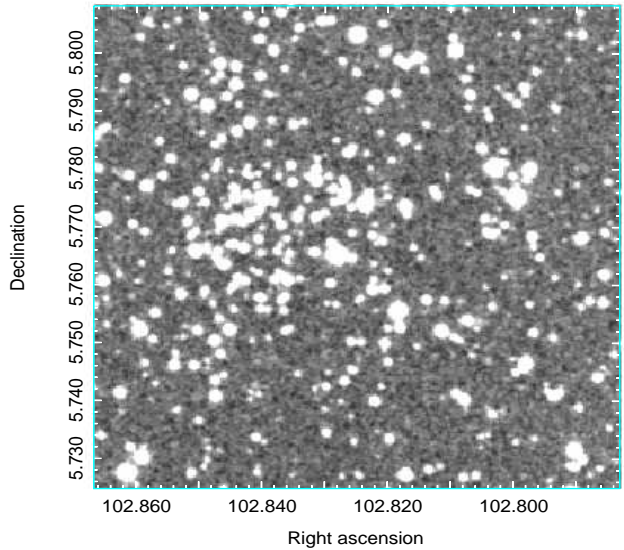}
\caption{Identification chart of Be 27 taken from the 
Digitzed Sky Survey (DSS).}
\label{id}
\end{center}
\end{figure*}

Be 27 ($\alpha_{2000} = 06^{h}51^{m}20.88^{s}$,
$\delta_{2000}=+05^{\circ} 46^{\prime} 19.2^{\prime\prime}$;
$l$=207$^\circ$.781, $b$=2$^\circ$.609 Cantat-Gaudin et al. 2018)
is an intermediate-age open cluster. 
The cluster is located in the third Galactic quadrant 
towards the Galactic anti-center region. There exist a handful
of previous studies of this cluster 
(Setteducati \& Weaver 1962; Hasegawa et al. 2004; Carraro \& Costa 2007; 
Donati et al. 2012). 
Hasegawa et al. (2004) have mentioned Be 27 as Biurakan 11 
and found an age = 2 Gyr and Z$_{metal}$ = 0.03 for this cluster. 
Carraro \& Costa (2007) reported the cluster's age around 2 Gyr
by fitting a solar metallicity isochrone.
Donati et al. (2012) report an age between 1.5 to 1.7 Gyr for Be 27.
As Donati et al. (2012) mentioned, Be 27 does not have 
a clear red giant branch and clump which causes some uncertainty 
in the parameters for this cluster. 
In such a scenario, it could be fruitful to fit a theoretical isochrone 
to a color-magnitude diagram (CMD) that has the field stars removed. 
We use the precise PMs from Gaia-DR3 to determine the membership status 
of the stars. The identification chart in 5$\times$5 arcmin$^2$ area
for Be 27 shown in Fig.~\ref{id}
which is taken from the STScI Digitized Sky 
Survey\footnote{https://archive.stsci.edu/cgi-bin/dss\_form?}.

Details regarding the Gaia data used in this work are provided 
in Section~\ref{OBS}. The field star decontamination, 
membership probability determination and
analysis of the associated results are presented in Section~\ref{PM}.
Using the most probable cluster members, some fundamental
parameters of Be 27 are calculated in Section~\ref{FUNDA}.\
In Section~\ref{ORB}, we describe the orbital picture of the cluster.
The conclusions of this work are presented in Section~\ref{CON}.

%%%%%%%%%%%%%%%%
%%%%%%%%%%%%%%%%

\section{Data}
\label{OBS}
%%%%%%%%%%%%%%%%
\begin{figure*}
\begin{center}
\includegraphics[width=9.5cm, height=9.5cm]{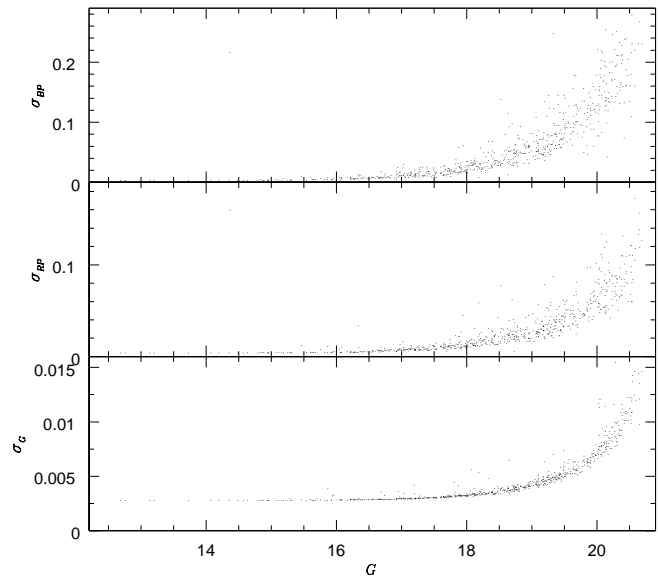}
\caption{Photometric errors in Gaia bands ($G$, $BP$ and $RP$) with $G$ magnitudes.}
\label{error_mag}
\end{center}
\end{figure*}
%%%%%%%%%%%%%%%%

\begin{figure*}
\begin{center}
\includegraphics[width=9.5cm, height=9.5cm]{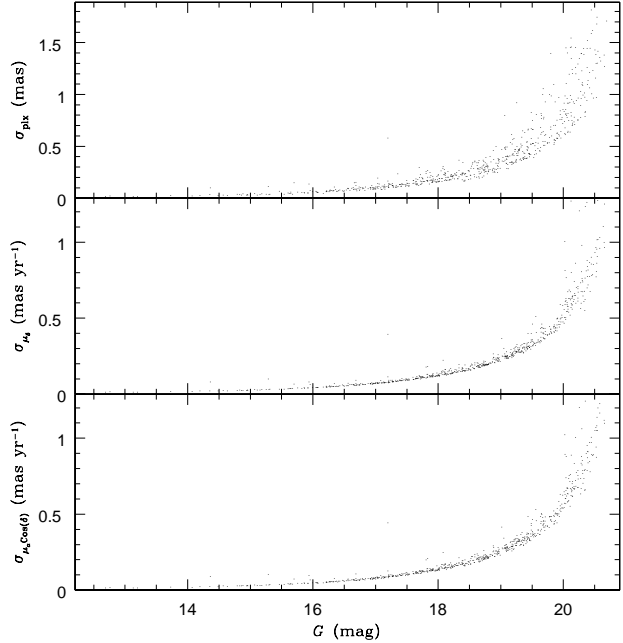}
\caption{The errors in parallax and PM components
are plotted here against Gaia $G$ magnitude.}
\label{error_pm}
\end{center}
\end{figure*}

We used the third data release from the Gaia satellite (Gaia Collaboration et al. 2016a,b, 2021). Gaia-DR3 contains the information about
positions ($\alpha, \delta$), parallaxes and PMs ($\mu_{\alpha} cos\delta , \mu_{\delta}$). The data for radial velocities also exist in the Gaia-DR3.
The photometry in Gaia data is available in three pass-bands: the white-light $G$, the blue $BP$, and the red $RP$ bands. We show the photometric errors
for these photometric bands with $G$ mag in Fig \ref{error_mag}. In Fig.~\ref{error_pm}, we plot the errors in parallax and PMs with $G$ mag. The value of
the mean PM error for the studied stars is $\sim0.05$~mas~yr$^{-1}$ for $G<$17 mag. The error becomes $\sim0.2$~mas~yr$^{-1}$ for stars with $G<$20 mag.
The mean error in parallax is $\sim0.28$~mas for stars brighter than 20 $G$ mag.
The median errors in $G, BP$ and $RP$ are 
$\sim$0.003, $\sim$0.03 and $\sim$0.02 for $G<$20 mag.

%%%%%%%%%%%%%%%%
%%%%%%%%%%%%%%%%

\section{Cluster membership}
\label{PM}

\subsection{Vector point diagrams}
\label{VPD}

\begin{figure*}
\begin{center}
\includegraphics[width=10.5cm, height=10.5cm]{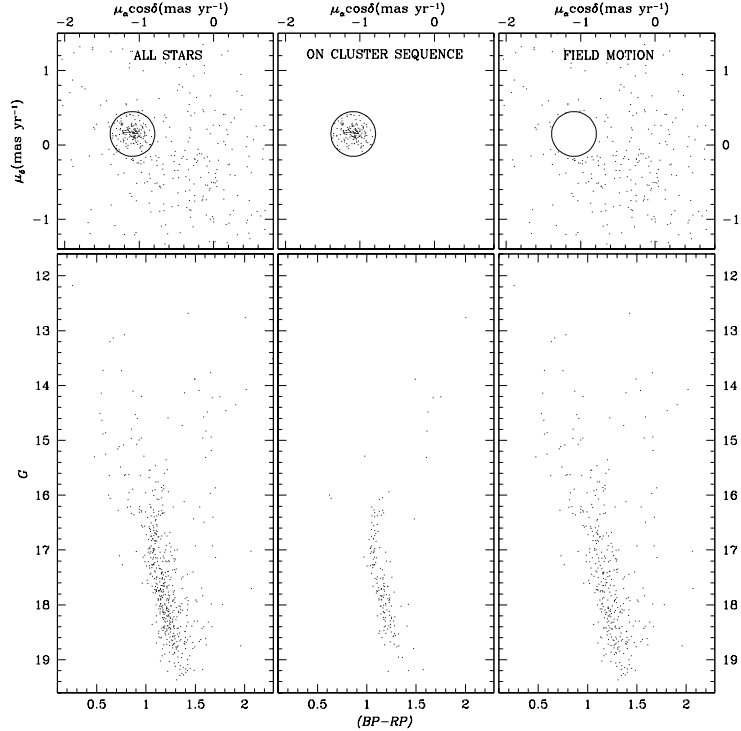}
\caption{(Top panels) The VPDs for the stars.
(Bottom panels) The ($(BP-RP)$, $G$) CMDs.
(Left) The entire sample of stars.
(Center) The stars presumed as cluster members with their PMs lying
within a radius of 0.3~ mas~ yr$^{-1}$ in the VPD.
(Right) The decontaminated field stars. }
\label{vpd}
\end{center}
\end{figure*}

The PMs can be used to distinguish cluster member stars
from the field stars in the same direction as the cluster. For this purpose, we initially plot between both components of the PMs
($\mu_{\alpha} cos{\delta}$, $\mu_{\delta}$). The resulting diagram is called the vector point diagram (VPD). Three separate VPDs for the stars in this work are shown in the top portions of Fig.~\ref{vpd}.
It is evident from the VPDs that the cluster's member stars tend to be located around a common point. The corresponding $G$ versus $(BP-RP)$ CMDs of Be 27 are shown in the lower portion of the same figure.
To make sure the inclusion of only the stars whose PMs are
measured more precisely by Gaia, we put a cut-off limit of
PM error $<$ 1 mas yr$^{-1}$ and parallax errors less than one mas.

The VPD has a circle of radius 0.3 mas yr$^{-1}$
which indicates our preliminary cluster membership criterion.
We also do not find a clear red giant branch and clump
in the CMD constructed in the middle section of Fig.~\ref{vpd}
as stated by Donati et al. (2012).

%%%%%%%%%%%%%%%%
%%%%%%%%%%%%%%%%

%%%%%%%%%%%%%%%%
\begin{figure*}
\begin{center}
\hbox{
\includegraphics[width=8.5cm, height=8.5cm]{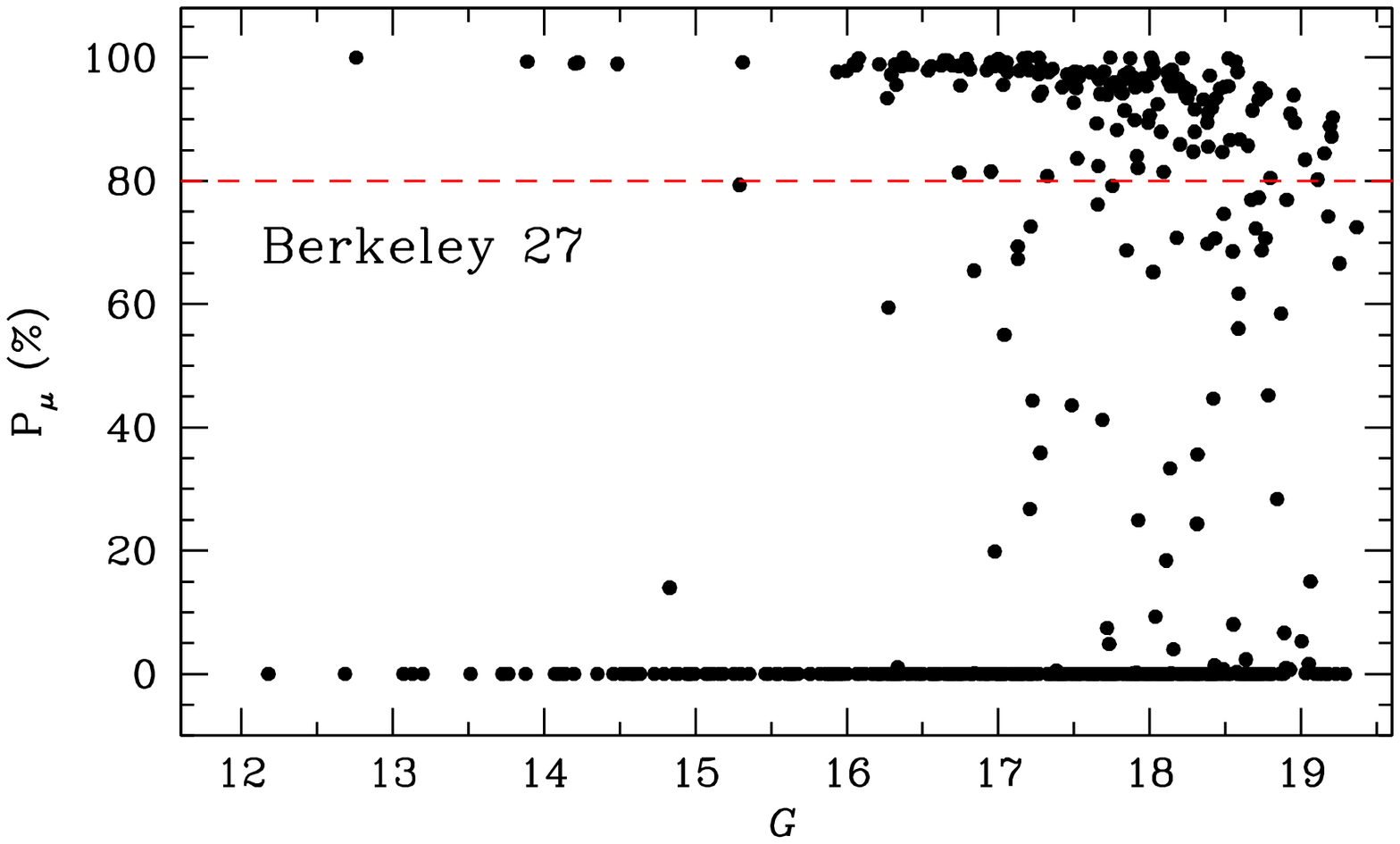}
\includegraphics[width=8.5cm, height=8.5cm]{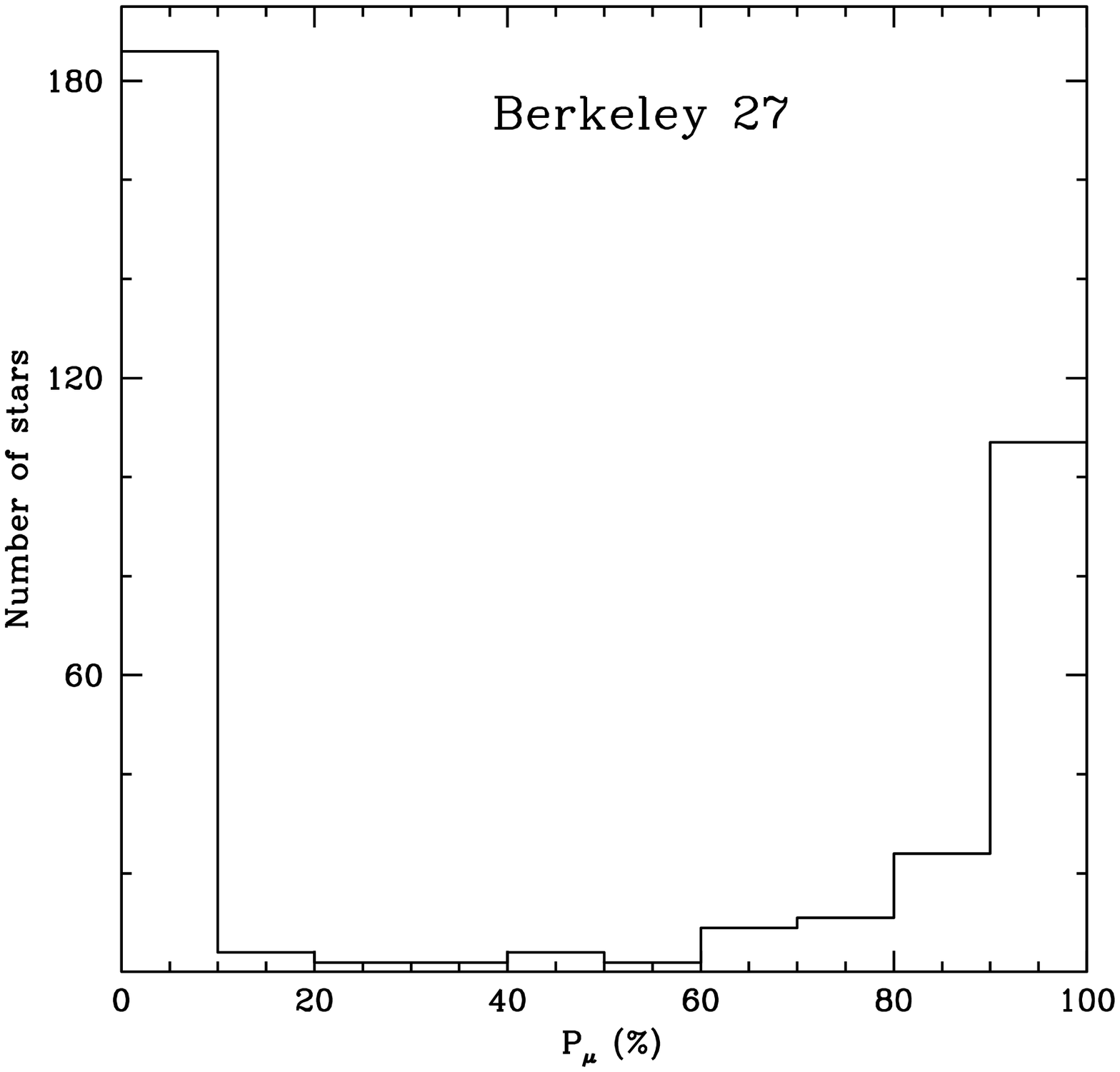}
}
\caption{(Left panel) Membership probability versus $G$ magnitude.
We also show here a horizontal dashed line for membership probability 80$\%$
above which we consider a star as the most probable cluster member.
(Right panel) Histogram of membership probability of the stars in
the field of Be 27.}
\label{mp_hist}
\end{center}
\end{figure*}
%%%%%%%%%%%%%%%%
%%%%%%%%%%%%%%%%%%%%%%%%%%%%%%%%%%%%%%%%%%%%%%%%%%%%%%%%%%%%%%%%%%%%%%%%%%
\begin{figure*}
\begin{center}
\includegraphics[width=9cm, height=9cm]{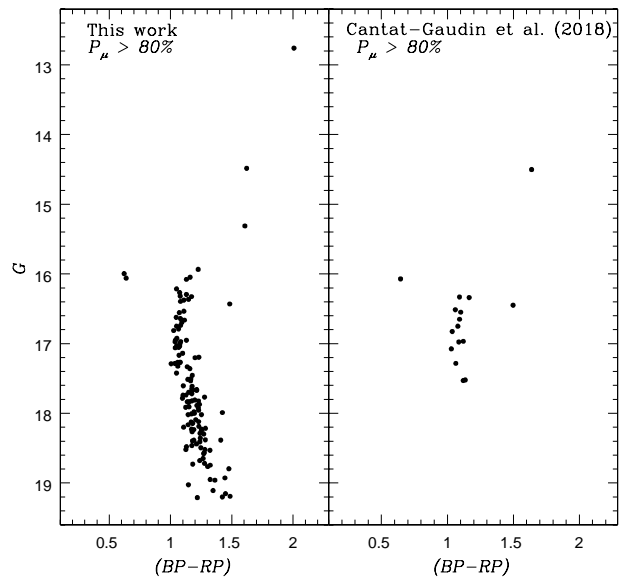}
\caption{Shown here are the CMDs of the most probable cluster
members of Be 27 from our analysis (left panel)
and from Cantat-Gaudin et al. (2018, right panel).}
\label{figmp90}
\end{center}
\end{figure*}
%%%%%%%%%%%%%%%%%%%%%%%%%%%%%%%%%%%%%%%%%%%%%%%%%%%%%%%%%%%%%%%%%%%%%%%%%%%

%%%%%%%%%%%%%%%%
\begin{figure*}
\begin{center}
\includegraphics[width=10.5cm, height=9.5cm]{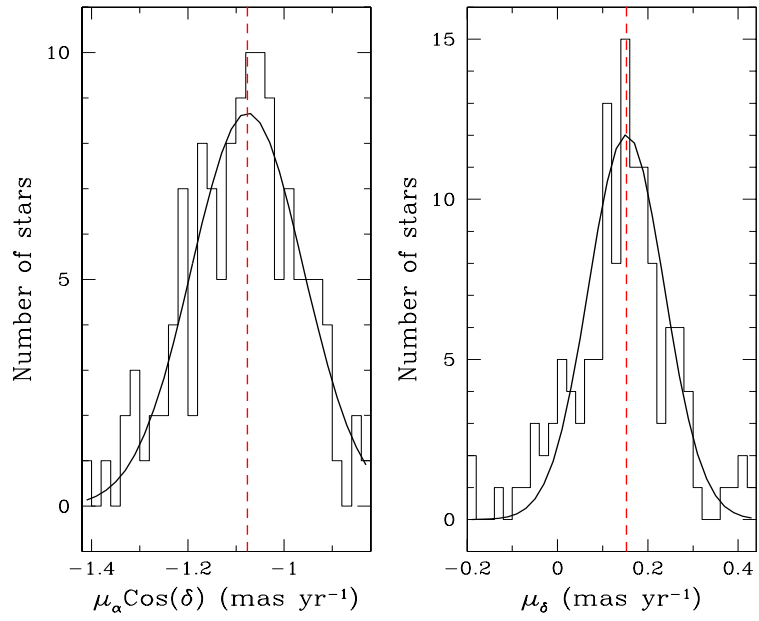}
\caption{ Histograms for the components of PM with a
Gaussian fitting done to determine the mean PM values.
The resulting mean values are shown by the  vertical dashed lines in both panels.
}
\label{pm_hist}
\end{center}
\end{figure*}
%%%%%%%%%%%%%%%%%%%%%%%%%%%%%%%%%%%%%%%%%%%%%%%%%%%%%%%%%%%%%%%%%%%%%%%%%%%
%%%%%%%%%%%%%%%%%%%%%%%%%%%%%%%%%%%%%%%%%%%%%%%%%%%%%%%%%%%%%%%%%%%%%%%%%%%
\begin{figure}
%\centering
\includegraphics[width=7cm, height=8cm]{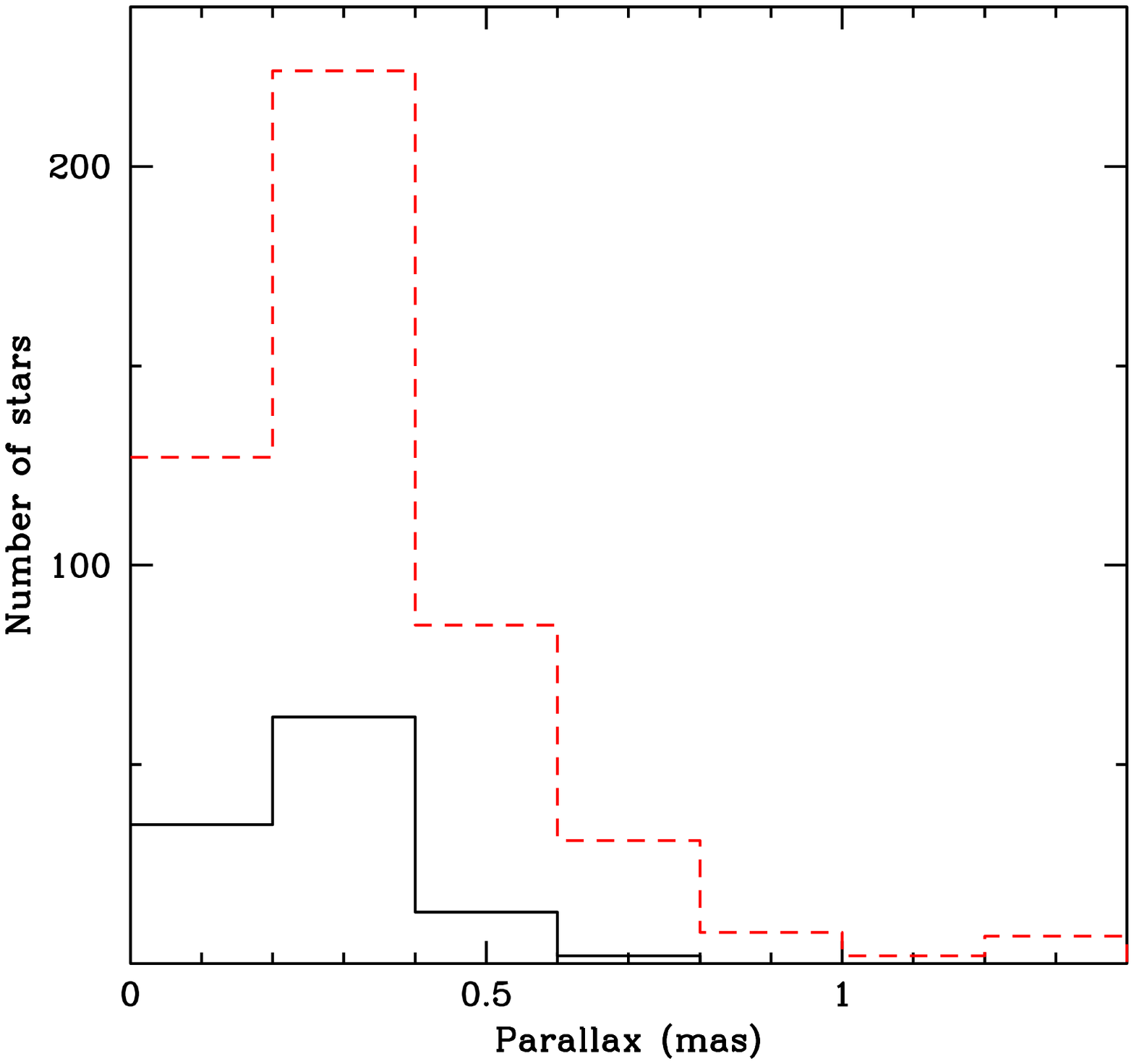}
\includegraphics[width=7.5cm, height=8.5cm]{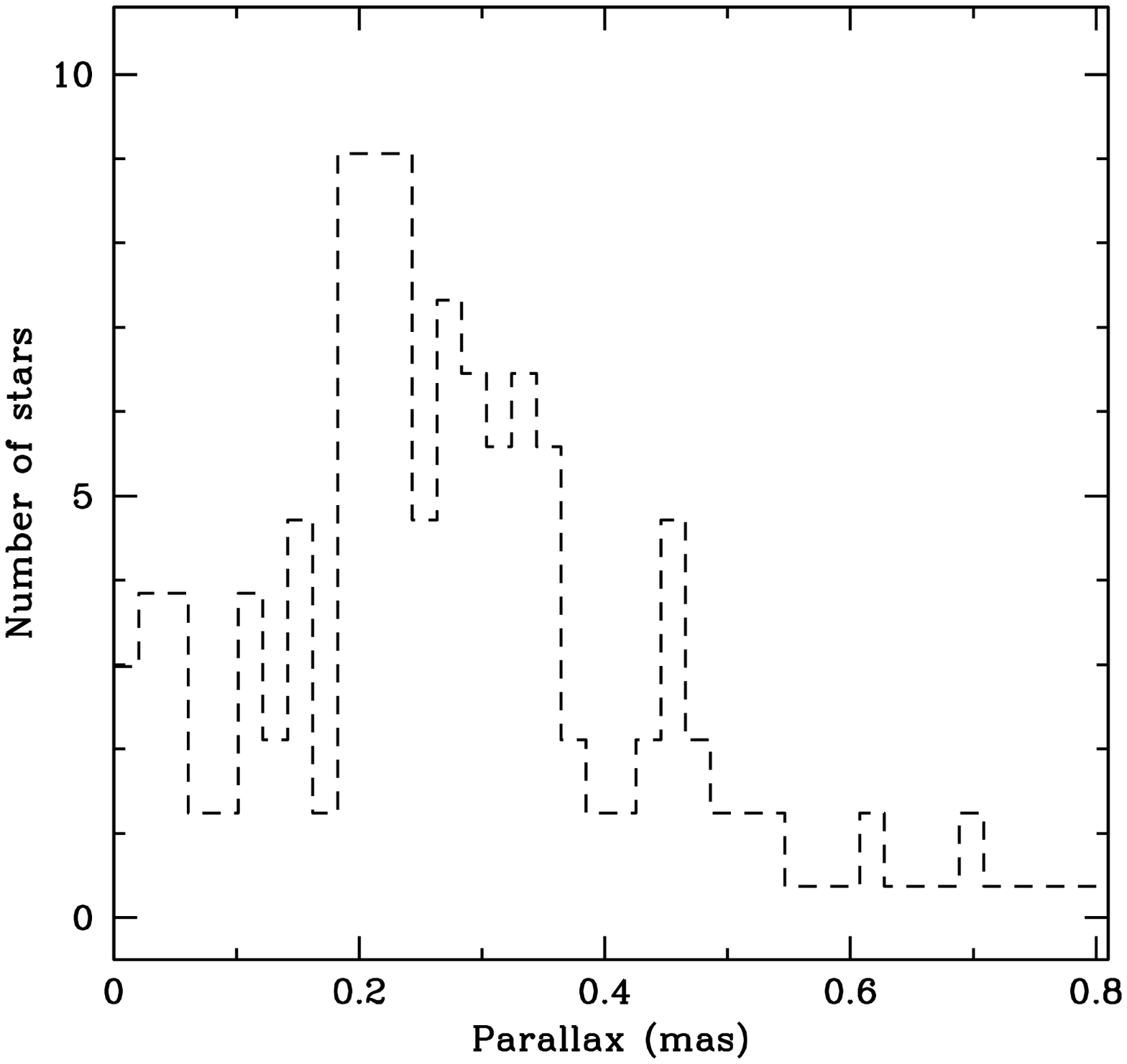}
\caption{
(Left) Histogram of Gaia-DR3 parallaxes for all stars in shown by
the red dashed line while the solid black histogram presents
the most probable cluster members.
(Right) For the most probable cluster members,
a Gaussian fit to the histogram provides the mean parallax value.
}
\label{figplx}
\end{figure}
%%%%%%%%%%%%%%%%%%%%%%%%%%%%%%%%%%%%%%%%%%%%%%%%%%%%%%%%%%%%%%%%%%%%%%%%%%%

%%%%%%%%%%%%%%%%%%%%%%%%%%%%%%%%%%%%%%%%%%%%%%%%%%%%%%%%%%%%%%%%%%%%%%%%%%%
\begin{figure*}
\begin{center}
\includegraphics[width=9.5cm, height=10.0cm]{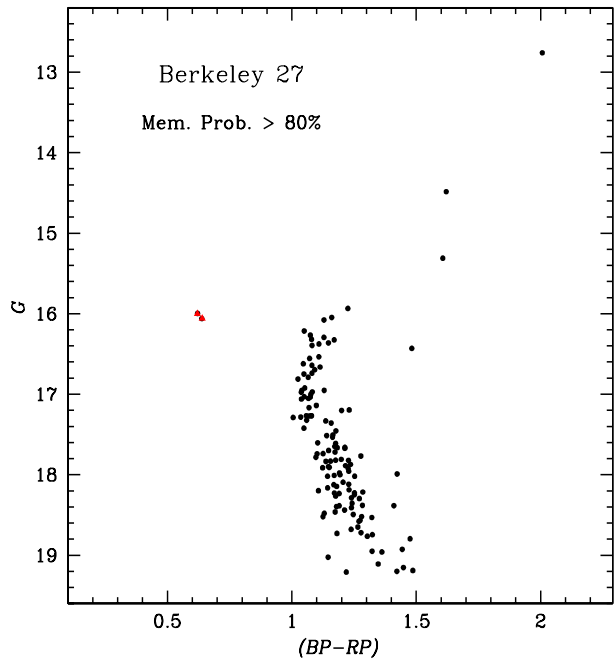}
\caption{
The position of two BSS in the CMD of the
most probable members of Be 27 is shown here by red triangles.
}
\label{bssposition}
\end{center}
\end{figure*}
%%%%%%%%%%%%%%%%%%%%%%%%%%%%%%%%%%%%%%%%%%%%%%%%%%%%%%%%%%%%%%%%%%%%%%%%%%%

\subsection{Membership probabilities}
\label{MP}

In Section~\ref{VPD}, we presented a robust method to decontaminate
field stars. However, we can do more analysis with the precise PMs from Gaia-DR3. This advantage of Gaia PMs led us to determine the membership
probabilities for the stars lying in the observing direction of Be 27. We used the method presented by Balaguer-N\'{u}\~{n}ez et al. (1998) to determined
the membership probabilities. This method was previously used for open and globular clusters (Sariya et al. 2018, 2021a,b; Bisht et al. 2020, 2021, 2022).
Sariya et al. (2018) give a detailed description of this method.

We put a PM error cut-off of 1 mas~yr$^{-1}$ and another condition that the parallax error should be better than 1 mas while calculating the distribution
functions defined in this method. These distribution functions are noted by $\phi_c^{\nu}$ (cluster star distribution) and $\phi_f^{\nu}$ (field star distribution),
For the cluster stars, we found
$\mu_{xc}$=$-$1.08 mas~yr$^{-1}$, $\mu_{yc}$= 0.16 mas~yr$^{-1}$ .
The dispersion value in the PMs of the cluster population is estimated to be ($\sigma_c$) = $\sim$0.06 mas~yr$^{-1}$. For the field stars, we found: ($\mu_{xf}$, $\mu_{yf}$) = ($-$0.35, $-$0.29) mas yr$^{-1}$
and ($\sigma_{xf}$, $\sigma_{yf}$) = (0.57, 0.72) mas yr$^{-1}$.\\

The resulting membership probabilities ($P_{\mu}$)
are plotted with Gaia's $G$ magnitude
in the left panel of Fig.~\ref{mp_hist}.
In the right panel of this figure, we show a histogram
of membership probabilities.
Using the distribution in histogram, we decided to use stars with
$P_{\mu}>$ 80\% as the most probable cluster members
for further analysis of Be 27.
In the left panel of Fig.~\ref{mp_hist}, a horizontal dashed line
is used to show this cut-off value of 80\%.

Conclusively, we obtained 131 stars with $P_{\mu}>$ 80\%
and lying within the limiting radius (3.74 arcmin, Section~\ref{RADIUS})
of Be 27.
We show these stars in the left panel of Fig.~\ref{figmp90}.
In the right panel of Fig.~\ref{figmp90}, we plot the CMD of
the stars with $P_{\mu}>$ 80\% according to the 
membership catalogue given by Cantat-Gaudin et al. (2018).
It is clear from these two CMDs that
as compared to Cantat-Gaudin et al. (2018),
we provide the most probable cluster members of Be 27
to a deeper magnitude ($G\sim$19.4 mag).

We used Gaussian function fitting to the
histograms of PM components
($\mu_{\alpha} cos{\delta}$ and $\mu_{\delta}$)
which is shown in Fig.~\ref{pm_hist}.
Thus, we obtained the values of mean PM of Be 27 =
($-1.076\pm0.008$, $0.152\pm0.007$) mas yr$^{-1}$.
These values are quite close to the values
($-$1.091, 0.147 mas yr$^{-1}$)
determined by Cantat-Gaudin et al. (2018)

The result for the parallaxes of Be 27 stars are shown
in Fig.~\ref{figplx}. The left panel of this figure
presents histograms of all the stars in our study
along with the histogram for the most probable cluster members.
A Gaussian fit to the histogram for the most probable cluster member is
shown in the right panel of Fig.~\ref{figplx}.
Our analysis gives us a value of 0.246 $\pm$ 0.011 mas
for the mean parallax of Be 27.

%%%%%%%%%%%%%%%%%%%%%%%%%%%%%%%%%%%%%%%%%%%%%%%%%%%%%
\subsubsection{The BSS of Be 27}

The presence of BSS in a stellar system is a symbol of the `delayed' evolution
for certain stars. 
The origin of this population of BSS can be attributed to the mechanisms
like mass transfer and merging due to collisions 
(Sandage 1953; McCrea 1964; Zinn \& Searle 1976; Hills \& Day 1976).

We detected two BSS as the most probable cluster members of Be 27.
Their position in the CMD is shown in Fig.~\ref{bssposition}.
Ferraro et al. (2012) defined three classes of BSS based on the
radial distribution of the BSS (see Sariya et al. 2021b).
The two BSS of Be 27 are located at a radial distance of
0.47 and 0.49 arcmin. Considering that both BSS are centrally located 
(within 0.5 arcmin), we can say that Be 27 belongs to the family III  
of the BSS' radial distribution according to the criterion given by 
Ferraro et al. (2012). A higher density of the stars in the 
central region could be the reason behind this scenario.

\section{Structural and fundamental parameters of Be 27}
\label{FUNDA}

%%%%%%%%%%%%%%%%%%%%%%%%%%%%%%%%%%%%%%%%%%%%%%%%%%%%%%%%%%%%%%%%%%%%%%%%%%%

\subsection{Radial density profile}
\label{RADIUS}

A radial density profile (RDP) is constructed in order to determine
the cluster's limiting radius.
The cluster's data is first divided into several concentric rings.
Then, for an $i^{th}$ zone, the stellar number density is calculated
as:
$R_{i}$ = $\frac{N_{i}}{A_{i}}$,
where $N_{i}$ is the number of stars in that zone
and $A_{i}$ is the surface area of it.
A King (1962) profile is fitted to the RDP.

The formula for the King profile is given below:

\begin{equation}
f(r) = f_{bg}+\frac{f_{0}}{1+(r/r_{c})^2}\\
\end{equation}

where $r_{c}$, $f_{0}$, and $f_{bg}$ are termed as
core radius, central density, and the background density level.
Fitting of the King profile to the RDP provides the
structural parameters for the studied cluster.
We calculated the following values of
$r_{c}$, $f_{0}$ and $f_{bg}$:
$0.56\pm0.12$ arcmin,
$33.43\pm9.36$ stars per arcmin$^{2}$
and $0.10\pm0.25$ stars per arcmin$^{2}$.

\begin{figure*}
\begin{center}
\includegraphics[width=7.5cm, height=7.5cm]{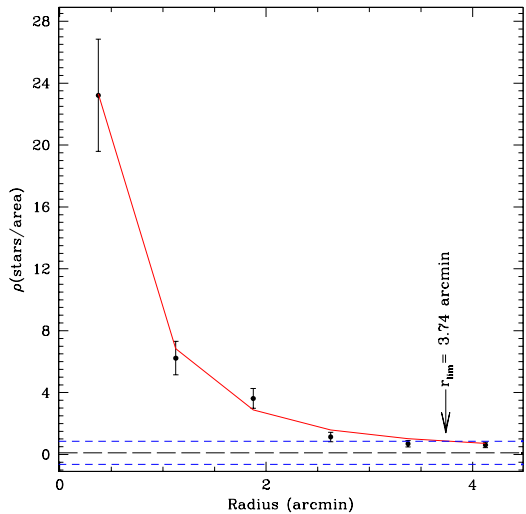}
\caption{The radial density profile for Be 27.
The King (1962) profile is shown by a curve.
The background density with 3$\sigma$ errors
is shown by the horizontal dashed lines.
}
\label{rdp}
\end{center}
\end{figure*}

The RDP along with the king profile is shown in Fig.~\ref{rdp}.
The background density along with the 3$\sigma$ errors in it
is also shown by the dashed lines in the figure.

In order to determine the limiting radius of Be 27, we used the
following equation mentioned by Bukowiecki et al. (2011):
$r_{lim}=r_{c}\sqrt(\frac{f_{0}}{3\sigma_{bg}}-1)$.
Thus, a limiting radius of 3.74$^{\prime}$ was obtained for Be 27.

%%%%%%%%%%%%%%%%
%%%%%%%%%%%%%%%%

\subsection{Age and distance}
\label{ISOCHRONE}

%%%%%%%%%%%%%%%%
\begin{figure*}
\begin{center}
\includegraphics[width=8.5cm, height=8.5cm]{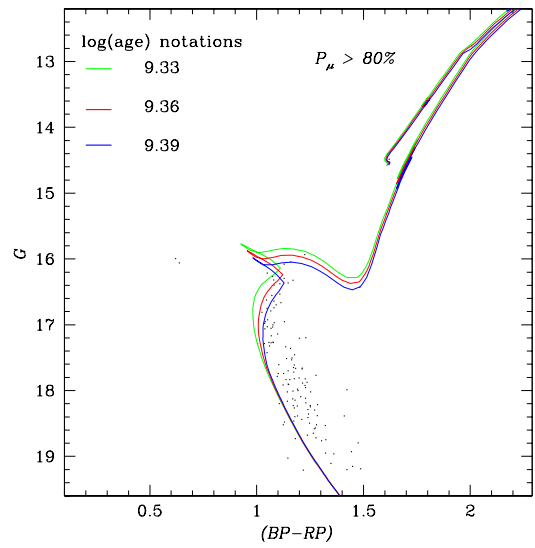}
\caption{The Gaia based CMD of the most probable cluster members
is matched with the theoretical isochrones provided by Bressan et al. (2012).
The fitted isochrones have a metallicity Z$_{metal}$ = 0.008 
and log(age) = 9.33$\pm$0.03.
}
\label{age}
\end{center}
\end{figure*}
%%%%%%%%%%%%%%%%

On the observed CMD of the most probable members of Be 27,
we fitted theoretical isochrones provided by Bressan et al. (2012)
to calculate the metallicity (Z$_{metal}$), age, heliocentric distance and reddening
of Be 27.
The isochrones' fitting to the CMD is shown in Fig.~\ref{age}.
These isochrones have a metallicity Z$_{metal}$=0.008 and
log(age)=9.33, 9.36 and 9.39.
Thus, the age of Be 27 is calculated to be 2.29$\pm$0.15 Gyr.
Also, the distance modulus ($m-M_{G}$) was derived as 14.26.
Upon using the equations mentioned by Hendy (2018),
we found the heliocentric distance of Be 27 = 4.8$\pm$0.2 kpc.
Donati et al. (2012) used several isochrones of different ages and metallicities.
Among these, they also used Padova models with Z$_{metal}$=0.008
which agrees with our results.
Kharchenko et al. (2013) catalogue lists the log(age) and distance values of
this cluster as 9.30 and 5.042 kpc.
These literature values agree with our results within the given errors.

\section{The cluster's orbit in the Galaxy}
\label{ORB}

Studying orbits is advantageous for understanding 
the stars, clusters, and Galaxies' formation and evolution processes. 
We utilized the method of Allen \& Santillan (1991) to acquire 
the Galactic orbits of Be 27. 
Bajkova \& Bobylev (2016) and Bobylev et al. (2017) have refined
Galactic potential model parameters using the new observational data 
for the Galactocentric distance R $\sim$ 0 to 200 kpc. 
The equations considered for the utilized models are 
described by Rangwal et al. (2019). 
The main fundamental parameters, namely 
the cluster center ($\alpha$, $\delta$ taken from Cantat-Gaudin et al. 2018), 
mean PMs ($\mu_{\alpha}cos\delta$, $\mu_{\delta}$), 
mean parallax, cluster's age and heliocentric distance ($d_{\odot}$)
have been used to define the orbital parameters in Be 27.

We converted equatorial space and velocity components 
into Galactic-space velocity components. 
The Galactic center is considered at 
($17^{h}45^{m}32^{s}.224$, $-28^{\circ}56^{\prime}10^{\prime\prime}$) 
and the North-Galactic pole is considered at 
($12^{h}51^{m}26^{s}.282$, $27^{\circ}7^{\prime}42^{\prime\prime}.01$,
Reid \& Brunthaler, 2004). 
To apply a correction for Standard Solar Motion 
and Motion of the Local Standard of Rest (LSR),
we employed the position coordinates of the Sun as ($8.3,0,0.02$) kpc 
and its velocity components as 
($11.1, 12.24, 7.25$) km~s$^{-1}$ (Schonrich et al. 2010). 
The transformed parameters in Galactocentric coordinate system 
are listed in Table \ref{inp}.

%%%%%%%%%%%%%%%%%%%%%%%%%%%%%%%%%%%%%%%%%%%%%%%%%%%%%%%%%%%%%%%%%%%%%%%%%%%%%%%
\begin{table*}
\centering
\small
\caption{Position and velocity components in the 
Galactocentric coordinate system. Here $R$ is the Galactocentric
distance, $Z$ is the vertical distance from the Galactic disc, 
$U$ $V$ $W$ are the radial tangential and the vertical
components of velocity respectively and $\phi$ is the position angle 
relative to the sun's direction.
}
\begin{tabular}{ccccccccc}
\hline\hline
Cluster   & $R$   &  $Z$  &  $U$   & $V$     & $W$    & $\phi$   \\
	  & (kpc) & (kpc) & (km~s$^{-1}$) &  (km~s$^{-1}$) & (km~s$^{-1}$) & (radian)    \\
\hline
Be 27 & 12.7665 & 0.24 & $-35.35\pm 4.67$  & $-236.19\pm 2.48$ &  $8.92\pm1.53$ & 0.18    \\
\hline
\end{tabular}
\label{inp}
\end{table*}
%%%%%%%%%%%%%%%%%%%%%%%%%%%%%%%%%%%%%%%%%%%%%%%%%%%%%%%%%%%%%%%%%%%%%%%%%%%%%%%

%%%%%%%%%%%%%%%%%%%%%%%%%%%%%%%%%%%%%%%%%%%%%%%%%%%%%%%%%%%%%%%%%%%%%%%%%%%%%%%
\begin{table*}
\centering
\small
\caption{Orbital parameters obtained using the Galactic potential model.
}
\begin{tabular}{ccccccccc}
\hline\hline
Cluster  & $e$  & $R_{a}$  & $R_{p}$ & $Z_{max}$ &  $E$ & $J_{z}$ & $T_{R}$ &$T_{Z}$  \\
	&    & (kpc) & (kpc) & (kpc) & (100 km~s$^{-1}$)$^{2}$ & (100 kpc km~s$^{-1}$) & (Myr) & (Myr) \\
\hline\hline
Be 27 & 0.0003  & 14.047 & 14.055 & 0.36 & $-$8.59  & $-$30.15 &  337.224  & 135.265 \\
\hline
\end{tabular}
\label{orpara}
\end{table*}
%%%%%%%%%%%%%%%%%%%%%%%%%%%%%%%%%%%%%%%%%%%%%%%%%%%%%%%%%%%%%%%%%%%%%%%%%%%%%%%

\begin{figure*}
\begin{center}
\hbox{
\includegraphics[width=5.6cm, height=5.6cm]{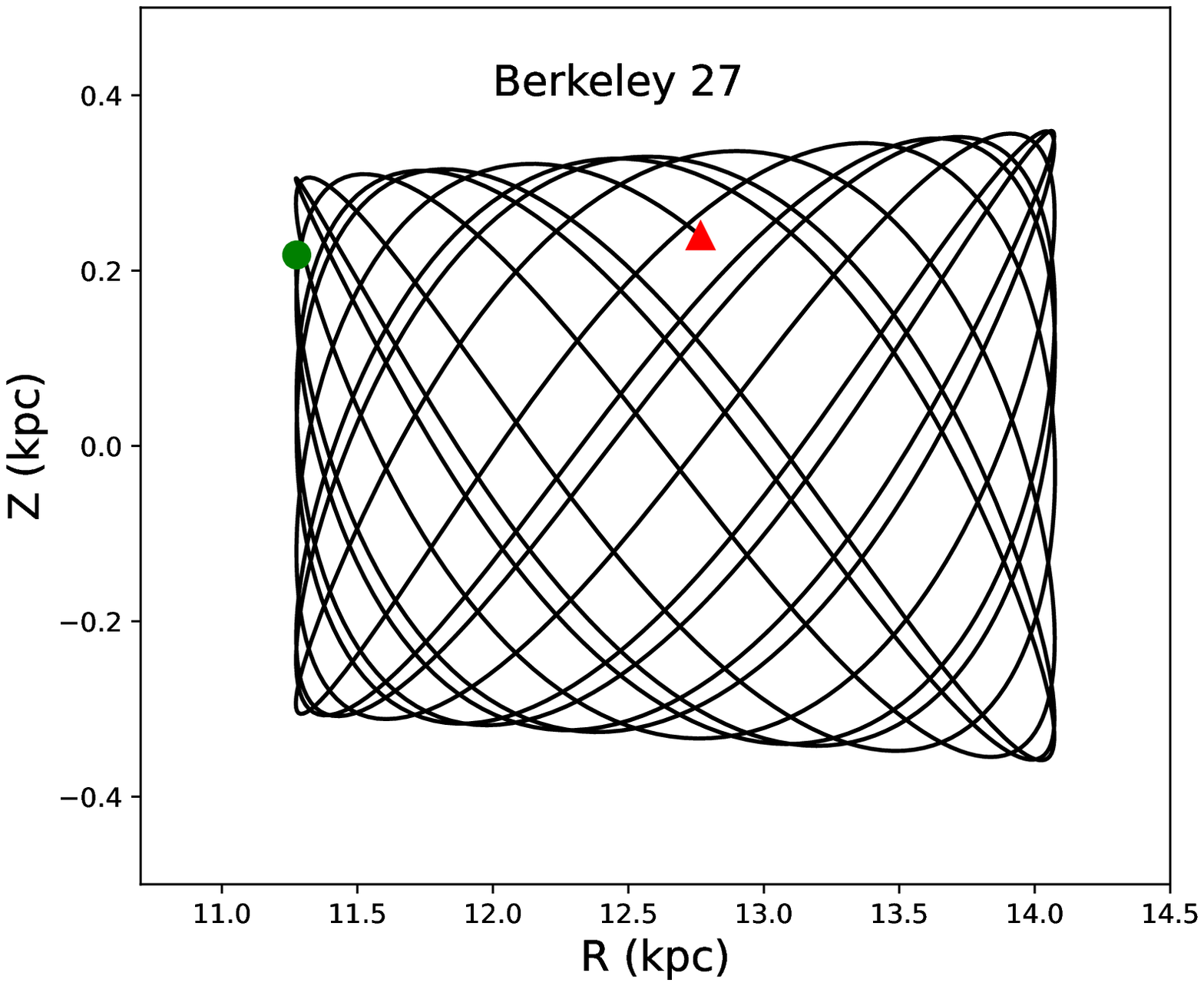}
\includegraphics[width=5.6cm, height=5.6cm]{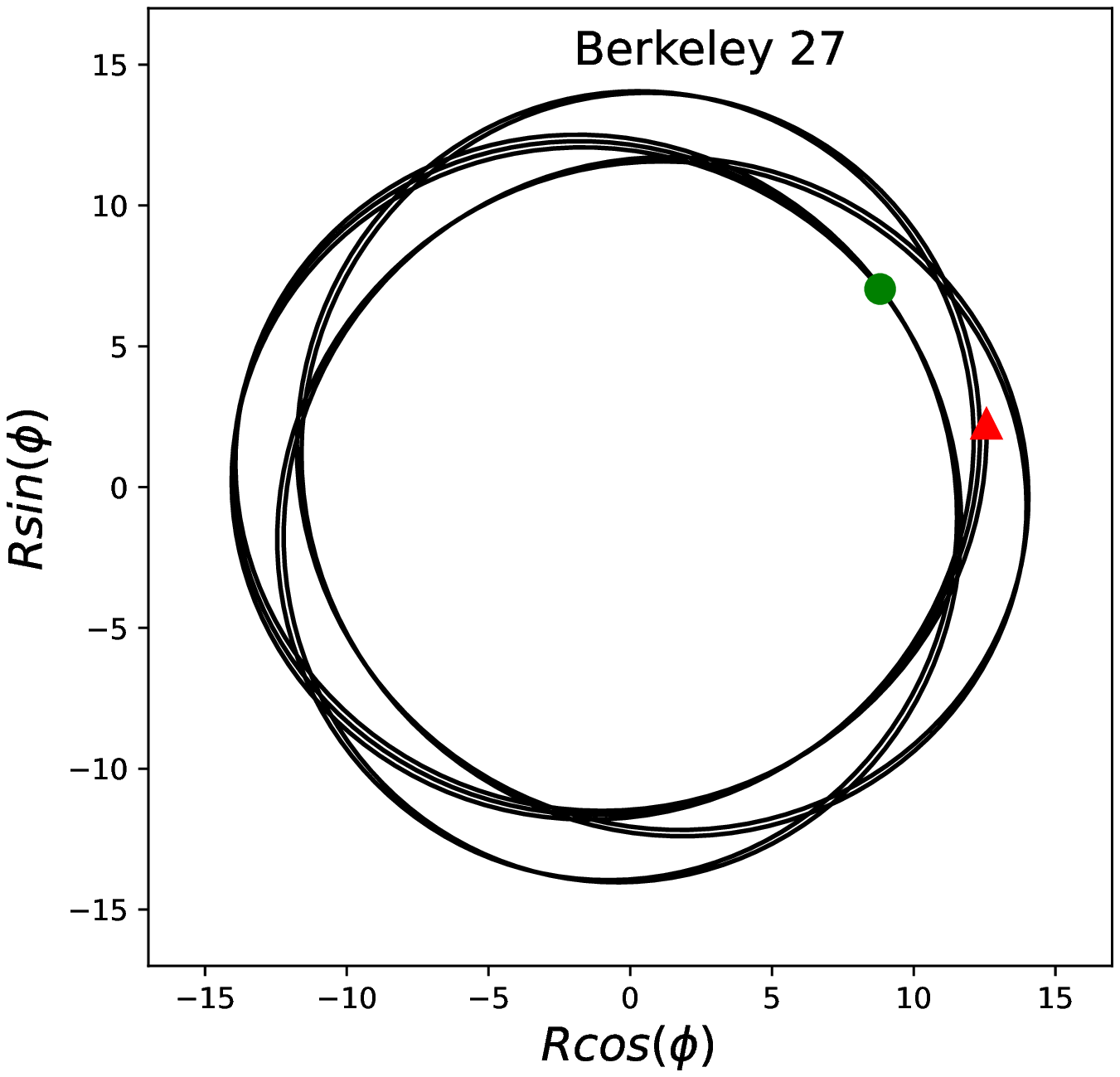}
\includegraphics[width=5.6cm, height=5.6cm]{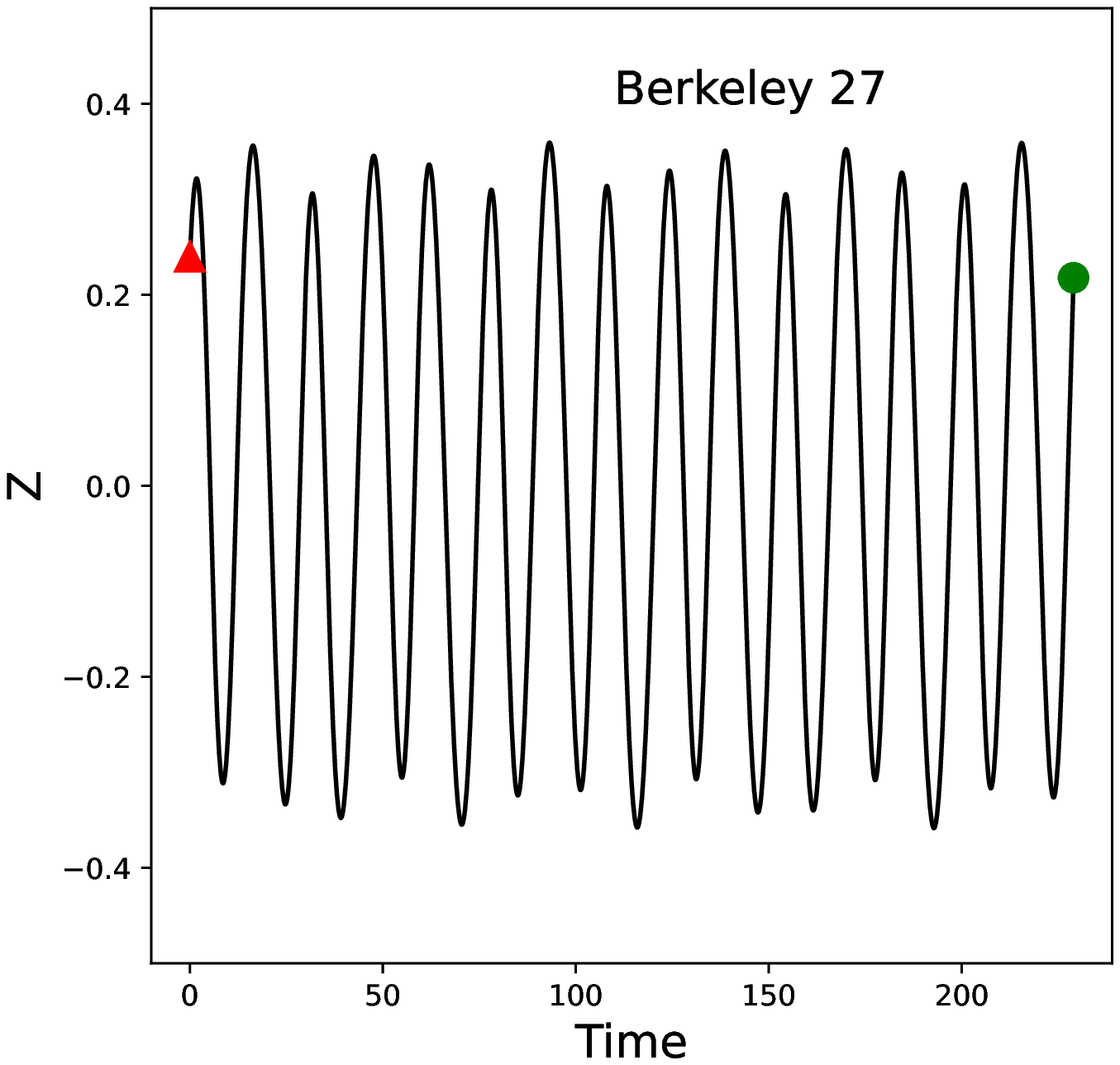}
}
\caption{Galactic orbits of the cluster Be 27 estimated with 
the Galactic potential model described in text in the time interval 
equal to the age of cluster. 
The left panel shows the side view and 
the middle panels show the top view of the orbits.
The right panels shows the motion of the cluster in the Galactic disc with time.
The filled circles and the triangles denote the birth and the present day 
positions of the cluster Be 27 
in the Galaxy.
}
\label{orbit}
\end{center}
\end{figure*}

Fig. \ref{orbit} shows the orbits of the cluster Be 27. 
The left panel of this figure indicates the cluster's motion 
in terms of distance from the Galactic center
and Galactic plane, which shows a two dimensional (2D) side view of the orbits. 
In the middle panel, the cluster motion is described in terms of 
$x$ and $y$ components of Galactocentric distance,
which shows a top view of orbits. 
The right panel of this figure indicates the cluster's motion 
in the Galactic disc with time. Be 27 follows a boxy pattern according to
our analysis. The eccentricity value we obtained is close to zero.
Hence, the studied cluster (Be 27) traces a circular path 
around the Galactic center. 
The birth and the present-day locations of Be~27 
in the Galaxy are represented by the filled circles 
and the triangles as shown in Fig. \ref{orbit}.
The various orbital parameters are thus obtained. 
These are listed in Table~\ref{orpara}. 
Here, $e$ denotes eccentricity, $R_{a}$ is the apogalactic distance, 
$R_{p}$ is perigalactic distance, $Z_{max}$ is the maximum distance 
traveled by cluster from Galactic disc, $E$ is the average energy of orbits, 
$J_{z}$ is $z$ component of angular momentum, $T_{R}$ 
is the time period of the revolution around the Galactic center, 
and $T_{Z}$ is the time period of vertical motion.

In these figures, we can see that the clusters' birth position
as well as the present day position is in the thick disc of the Galaxy. 
Hence, the cluster should have had a minimal interaction
with the Galactic thin disc. 
As a result, the cluster is not expected to be affected much 
by the Galactic tidal forces originating from thin disc. 
Also, Be 27 is not affected by the inner part of the Galaxy 
since it is orbiting outside the solar circle.

\section{Conclusions}
\label{CON}

We presented a study of intermediate-age open cluster
Be 27 using the Gaia-DR3 data.
Gaia PMs were used to determine the most probable cluster members.
The main conclusions of the current study are given below:

\begin{itemize}

\item The limiting radius of Be 27 is calculated as 3.74 arcmin
using the RDP.\

\item We identified 131 most probable cluster members
which are also lying within the limiting radius of the cluster.
The mean PM of Be 27 is determined as
($\mu_{\alpha} cos{\delta}$ and $\mu_{\delta}$)=
($-1.076\pm0.008$, $0.152\pm0.007$)~mas~yr$^{-1}$.

\item In the cluster's central area (within 0.5 arcmin)
2 most probable BSS are identified.\

\item Bressan et al. (2012) isochrones of metallicity Z$_{metal}$=0.008
were fitted to the CMD of most probable cluster members.
Thus, we determined the heliocentric distance of Be 27=$4.8\pm0.2$ kpc.
The cluster's log(age) is estimated to be $9.36\pm0.03$ .

\item The Galactic orbits and orbital parameters are evaluated 
for Be 27 using potential Galactic models. 
We found this cluster is orbiting in a boxy pattern 
outside the solar circle and traced the circular path 
around the center of the Galaxy.

\end{itemize}

\section{Acknowledgments}
\small{
The authors are grateful to the reviewer for useful remarks on this paper.
This work is supported by the grant from the
Ministry of Science and Technology (MOST), Taiwan.
The grant numbers are MOST 110-2112-M-007-035 and MOST 111-2112-M-007-035.
This work has made use of data from the European Space Agency (ESA) mission
Gaia (http://www.cosmos.esa.int/gaia), processed by the Gaia Data
Processing and Analysis Consortium (DPAC, http://www.cosmos.
esa.int/web/gaia/dpac/ consortium). Funding for the DPAC has been
provided by national institutions, in particular the institutions
participating in the Gaia Multilateral Agreement.
}

\section{Data Availability}

The Gaia-DR3 data used in this paper are publicly available.

\end{document}